\documentclass{iau} 
\usepackage{graphicx}

\title[Central stars of PNe] %% give here short title %%
{Central Stars of Planetary Nebulae}

\author[David Jones]   %% give here short author list %%
{David Jones$^{1,2}$
}

\affiliation{$^1$Instituto de Astrof\'isica de Canarias,\\ E-38205 La Laguna, Tenerife, Spain \\ email: {\tt djones@iac.es} \\[\affilskip]
$^2$Departamento de Astrof\'isica, Universidad de La Laguna,\\ E-38206 La Laguna, Tenerife, Spain}

\pubyear{2017}
\volume{323}  %% insert here IAU Symposium No.
\jname{Planetary nebulae: Multiwavelength probes of stellar and galactic evolution}
\editors{X. Liu, L. Stanghellini, \& A. Karakas, eds.}
\begin{document}

\maketitle

\begin{abstract}
In this brief invited review, I will attempt to summarise some of the key areas of interest in the study of central stars of planetary nebulae which (probably) won't be covered by other speakers' proceedings.  The main focus will, inevitably, be on the subject of multiplicity, with special emphasis on recent results regarding triple central star systems as well as wide binaries which avoid a common-envelope phase.  Furthermore, in light of the upcoming release of Kepler's Campaign 11 data, I will discuss a few of the prospects from that data including the unique possibility to detect merger products.

%% add here a maximum of 10 keywords, to be taken form the file <Keywords.txt>
\end{abstract}

\firstsection % if your document starts with a section,
              % remove some space above using this command.
\section{Triple star systems}
Recent works (e.g.\ Bear \& Soker, 2016; Akashi 2016) have sparked interest in the subject of higher order multiples at the heart of planetary nebulae (PNe) and the role they might play in forming the most unusual and/or asymmetrical PNe, particularly given the large fraction observed in solar-like main sequence stars (Raghavan \etal\ 2010).  However, to-date, there is no confirmed triple central star known.

\subsection{SuWt~2}
One of the best candidate triple central stars is NSV19992, the bright star near the apparent centre of the PN SuWt~2.  \cite{exter10} reported that this star is, in fact, a binary system consisting of two A-type main sequence stars with an orbital period of 4.9 days.  Neither of these stars could provide the ionising flux for the nebula, and as such hypothesise that the A-type binary may form part of a triple system along with the nebular progenitor.  Combining data from multiple sources, they find some evidence for shifts in the systemic velocity of the binary that would be consistent with a possible third component.  However, the morphology of SuWt~2 (Jones \etal\ 2010) is that of a rather canonical bipolar, seemingly inconsistent with the predicted ``messy'' structures for triple evolution (Bear \& Soker 2016).  In order to probe the possible triple system, \cite{jones16} acquired high-resolution VLT-UVES spectra of NSV19992 over the course of one-year, finding that there was no appreciable shift in the systemic velocity over the entire observing period.  Furthermore, with the binary parameters ($\gamma$, $K_1$, $K_2$, $T_0$) they derive, much of the data from \cite{exter10} phases well, with no need for a shift in systemic velocity (Jones \& Boffin 2016).  Collectively, this is strongly indicative that the A-type binary does not for a triple system with the nebular progenitor.  Additionally, the as the systemic velocities of the nebula and A-type binary are found to differ by more than 15 km s$^{-1}$, it seems unlikely that the two are in any way connected and rather that the A-type binary is just a field star which, by chance, lies close to the projected centre of the nebula.

\subsection{Other chance alignments}
Somewhat surprisingly, \cite{boffin16} also find that the bright star near the projected centre of PN M3-2 is a binary system comprised of two A-type main sequence stars, this time with a $\sim$2 day period.  The chances of finding a chance alignment of a double A-type binary and a PN is slim (but not impossible), the probability of finding two such chance alignments is, however, exceptionally small.  This might perhaps be taken as an indication that there may be some connection between the systems, for example, that one of the components of each binary system might be a born-again star.  For SuWt~2, this interpretation seems extremely unlikely given that the stellar parameters derived by \cite{exter10} are not consistent with this evolution.  Similarly for M3-2, a faint blue star, which is most likely the true central star, can be seen in deep imagery with good seeing (Boffin \etal\ 2016).

While not an A-type binary, it is worth to mention that \cite{mendez16} recently showed that the bright A-type star, believed to form one half of a binary system, at the heart of NGC~1514 is, in fact, unrelated to the nebula and just another chance alignment.  As such, chance alignments with field stars may not be as uncommon as predicted.

\section{Long- and intermediate-period binary central stars}
The connection between binary central stars and PN shaping is now beyond doubt (Hillwig \etal\ 2016).  Given that other authors (e.g.\ Nordhaus 2016) will summarise the importance (and current state of our understanding) of post-common-envelope central stars as part of these proceedings, I will attempt to cover the longer-period systems, summarising our current knowledge of long- and intermediate-period binary central stars.

\begin{figure}[b]
% \vspace*{-2.0 cm}
\begin{center}
 \includegraphics[width=0.9\textwidth]{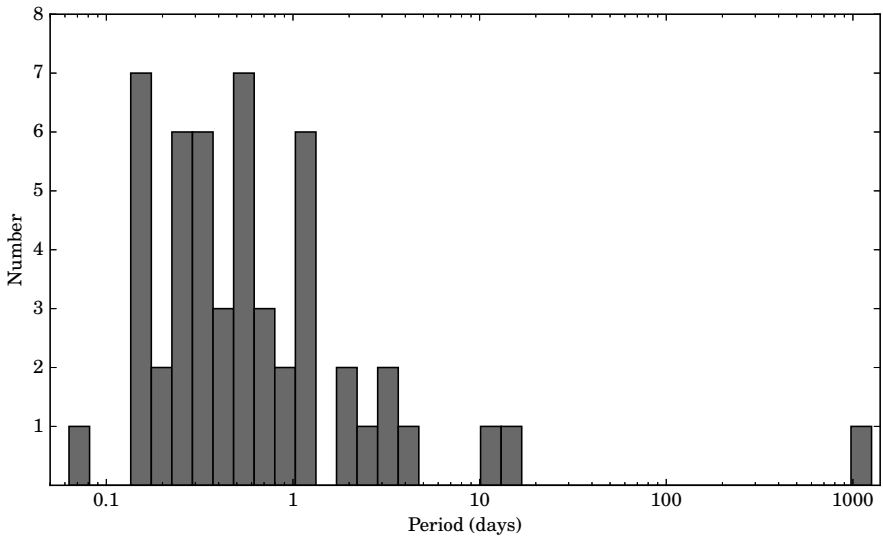} 
% \vspace*{-1.0 cm}
 \caption{Period distribution of known binary central stars of PNe (data from the list maintained at http://drdjones.net/bCSPN).}
   \label{fig:periods}
\end{center}
\end{figure}

With more than 50 known binary central stars (see figure \ref{fig:periods}) only a handful are known with periods greater than a few days.  I will categorise these systems by the technique used in their discovery, and try to explain the inherent difficulties (and positive points) for each technique showing the difficulty faced in unveiling the hidden (but clearly very important) population of wide binaries in PNe.

\subsection{Radial velocity monitoring}

While the amplitude of photometric variability, due to irradiation or ellipsoidal modulation, falls off rapidly with longer periods, radial velocity variables (at favourable inclinations) remain detectable with modest resolution spectrographs out to periods of several days (up to several hundred days for high resolution spectrographs).  However, this technique does have severe limitations in that it is intrinsically limited to brighter central stars (as high signal-to-noise is essential) and extremely costly in terms of observing time (as several different processes can contribute to radial velocity variability, it is essential to observe exhaustively in order to establish the periodic component associated with orbital motion from other sources of variability, for example, wind variability, De Marco \etal\ 2004).

Recently, radial velocity monitoring has been used to great effect by \cite{vanwinckel14} in discovering the wide binaries at the heart of BD+33$^{\circ}$2642 and LoTr~5 (which, coincidentally, was previously believed to be a triple system, a hypothesis not consistent with the observations of Van Winckel \etal{}) and by \cite{manick15} in the discovery of the 4 day period, post-common-envelope, Wolf-Rayet binary in NGC5189.

\subsection{Giant and/or chemically polluted secondaries}

Several systems are known where the star at the apparent centre of the nebula cannot have produced the nebula but, as previously shown, this cannot be taken as evidence that the observed star forms part of a binary (or higher-order multiple) with the nebula progenitor (M\'endez \etal\ 2015).  There are cases, however, where the star at the apparent nebular centre shows evidence of chemical pollution with AGB or post-AGB material (s-process elements, for example), this can be considered a strong indication that the star is associated with the nebula and that its progenitor was the source of chemical pollution (Miszalski \etal\ 2013).  In these cases, while the radius of the polluted star and the level of pollution can place limits on possible periods, further radial velocity study is essential to constrain precisely the true period of the binary (e.g.\ LoTr~5 was known to host a G-type giant star before it's orbital period was found by Van Winckel \etal\ 2014).  As the periods here are usually $\sim$100 days, period determination therefore requires long-term monitoring with intermediate- to high-resolution spectrographs, which is often particularly difficult given the competitive nature of telescope time allocations.

\subsection{Infrared excesses}

The search for infrared excesses (Barker 2016), where the SED of the hot primary shows an additional component (often peaking in the red or infrared) due to a cooler companion, is perhaps the most promising methodology for revealing binary central stars as the technique does not depend on periodicity or inclination.  Similarly, it is sensitive to a wide range of secondary masses (much more so that any radial velocity monitoring or photometric variability monitoring).  However, the technique is fraught with difficulty, requiring extremely high precision photometry in multiple bands often where the nebular contamination is appreciable and variable with wavelength, and unfortunately offers no clue as to the possible period of the binary (and, as such, central stars displaying IR excesses might be most appropriately used as a pre-selector for follow-up with other photometric or spectroscopic techniques, Aller 2016).  Interestingly, the results from surveys for IR excess (De Marco \etal\ 2013, Douchin \etal\ 2015) indicate an extremely significant total (both close and wide) binary fraction further reinforcing that a binary pathway is responsibe for a great number of PNe, perhaps even crucial in a majority of cases.

\section{Merger products}

Finally, it is important to mention that mergers with stellar or sub-stellar companions may also play an important role in the formation of aspherical planetary nebula (Nordhaus \& Blackman 2006).  Perhaps the main reason that mergers are so frequently over-looked is the extreme difficulty in their detection.  However, with the advent of \textit{Kepler} and world-wide photometric campaigns (e.g.\ The Whole Earth Telescope), the detection of merger products is now within reach, with the first such object being reported by \cite{handler13} and \cite{demarco15} in NGC~6826.  With another $\sim$200 PNe to be observed by \textit{Kepler} as part of Campaign 11 (Jacoby \etal\ 2016) and several ground-based photometric campaigns underway (Sowicka \etal\ in preparation), we may soon be in a position to observationally constrain the importance of a merger pathway for PN formation.

%\begin{discussion}
%\discuss{Helge Todt}{How can you distinguish between excess from hot dust and that from a binary companion?}
%\discuss{Rodolfo Montez}{Have you studied LoTr5 which has also been suggested to host a triple?}
%\discuss{Hans Van Winckel}{Is is possible that the A-stars are born again?}
%\end{discussion}

\end{document}